\documentclass[doublecol]{epl2} 

\usepackage{graphicx}  
\usepackage{dcolumn}   
\usepackage{bm}        
\usepackage{amssymb}   
\usepackage{booktabs}
\usepackage{enumitem}
\usepackage{calc}
\usepackage[utf8]{inputenc}
\usepackage[intlimits]{amsmath}
\usepackage{epsfig}
\usepackage{rotating}
\usepackage{transparent}
\usepackage{amsfonts}

\newcommand{\Var}{\mathbb V\mathrm{ar}}

\begin{document}
\title{Environmental control of microtubule-based bidirectional cargo transport}
\author{Sarah Klein\inst{1,2} \and C\'ecile Appert-Rolland\inst{1} \and Ludger Santen\inst{2}}

\institute{                    
  \inst{1}Laboratory of Theoretical Physics, CNRS (UMR 8627), University Paris-Sud
B\^atiment 210, F-91405 ORSAY Cedex, France\\
  \inst{2} Fachrichtung Theoretische Physik, Universit\"at des Saarlandes D-66123 Saarbr\"ucken, Germany}


\abstract{
Inside cells, various cargos are transported by teams of molecular motors. Intriguingly, the motors involved
generally have opposite pulling directions, and the resulting cargo dynamics is a biased stochastic motion.
It is an open question how the cell can control this
bias. Here we develop a model which takes explicitly into account the elastic coupling of the cargo with
each motor. We show that bias can be simply controlled
or even reversed in a counterintuitive manner via a change in the external force exerted on the cargo or a variation of the
environmental properties. 
Furthermore, the superdiffusive behavior found at short time scales indicates the emergence of motor cooperation induced by cargo-mediated coupling.
}

\maketitle

\section{Introduction}

In cells, most of the active transport processes, which are essential for cellular functions, 
are driven by molecular motors. These molecular motors are proteins having the 
ability to move preferentially in a defined direction on the polar filaments of the 
cytoskeleton \cite{alberts2002}. The three most well-known molecular 
motors' families involved in transport are myosins, which move on actin
filaments, dyneins and kinesins, which use microtubules (MT) as tracks
\cite{mallik_g2004}. Kinesin motors are stepping preferentially toward the growing
(or plus-) end of MTs while dynein motors walk in the opposite direction. 

Molecular motors can step individually or transport cargos along the cytoskeletal
filaments. In order to generate forces large enough \cite{Chowdhury2013} to move a big
cargo in the crowded environment of the cell, cargos are often transported by teams of 
molecular motors \cite{kural2005}. This is obviously beneficial if motors of the same
type are attached to the cargo, since the force can be
distributed between them. It enhances the processivity of the cargo and its ability 
to resist against forces emerging when transporting the cargo.

In many cases, however, motors that are attached to a given cargo
pull in opposite preferential directions. Surprisingly, the attachment of two kinds of motors is not only observed
for objects like mitochondria, which have to
be spread out in the whole cell volume 
\cite{hollenbeck_s2005}, but also for cargos which have a definite target, for example
to be transported from the cell center to the
membrane or {\it vice versa} \cite{welte2004}. 

Although the attachment of two kinds of motors should enable bidirectional transport
it is expected that, due to the difference in the characteristics of the various types of motors
attached to the cargo \cite{kunwar2011} and possibly in the number of attached motors,  the 
cargo undergoes a {\it biased} stochastic walk if the MT network is oriented. 

Depending on the given motor-cargo system and the environment of the filaments different types of motion have been observed, which can be controlled 
by different mechanisms.

Some pigment cells (melanosomes) for example have the ability to switch
between two states, in which the pigments are either dispersed or
aggregated at one extremity of the
filaments~\cite{gross2002b,nascimento_r_g2003}.
The mechanisms that allow for such a transition from a
non-biased to a biased motion are not yet well understood but 
have been related to signaling processes which regulate the 
activity of the attached motors \cite{nilsson2013}. 

Next to the active regulation of the motor dynamics, for example in cell signaling processes, also the cellular environment plays an important role. 
Recent {\em in vivo} experimental studies on cargos
transported bidirectionally have revealed a very
complex dynamical behavior.
Super- as well as subdiffusive regimes 
of the cargos' trajectories have been identified \cite{kulic2008,caspi_g_e2002,robert2010}. 
Bidirectional cargo transport on 
MT networks has also been studied {\it in vitro} where enhanced diffusion was obtained as well \cite{salman2002}. \\
It was also observed that a change in direction often occurs 
due to cellular obstacles like other cargos or the cytoskeleton 
itself \cite{Bruno2008}. In order to disentangle the action of different external control mechanisms on motor-cargo 
complexes we study their dynamics by means of a theoretical model. 

Several approaches have been proposed in the past years to describe cargo transport mediated by one \cite{kunwar2008,korn2009,Berger2012} or two types of motors \cite{kunwar2011,mueller_k_l2008,Zhang2013}. The stochastic model we introduce in this work which is similar to the one in \cite{kunwar2011} describes bidirectional cargo-motion along microtubles driven by teams of kinesin and dynein motors. We will show that a variation of
the environment changes the cargo-dynamics in a non-trivial way and can even invert the direction of the bias. 
A non-trivial response to the effective viscosity of the environment (representing the crowdedness of the surrounding cytosceleton) is also obtained as a result of the complex
interaction between the attached molecular motors.
Additionally, we show that, as observed in several \textit{in vivo}
experiments \cite{caspi_g_e2002,robert2010},
super-diffusive particle motion can be
observed in our model.

\section{Model}\label{Model}
\begin{figure}[bt]
\centering{\includegraphics[width = 0.35\textwidth]{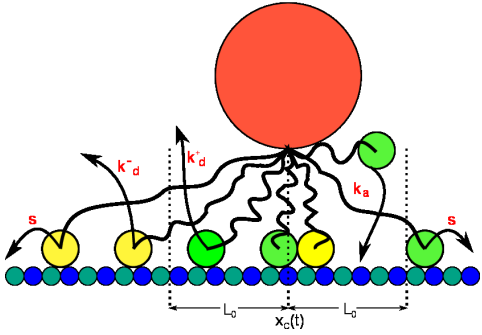}}
\caption{Sketch of the model dynamics. The cargo (red) is pulled by two families of molecular motors (yellow and green) hopping in opposite directions on the MT.} \label{skizze}
\end{figure}
We introduce a stochastic model for the transport of a cargo by teams of molecular motors along a single microtubule. 
In our model, which is inspired by the bidirectional cargo transport models of \cite{mueller_k_l2008,kunwar2011},
 $N_+$ and $N_-$ motors are tightly bound to the cargo and pull it in plus- and minus-direction, respectively. Throughout the paper we shall take $N_+=N_-=5$ if not stated otherwise.
 In contrast to \cite{mueller_k_l2008} we take every single motor position $x_i$ into account and calculate the thereby generated force $F_i$ on the cargo as in \cite{kunwar2011}. We model the tail of the molecular motors, which permanently connects the motor domain with the cargo, as a linear spring with untensioned length $L_0$ and spring constant $\alpha$ (Fig. \ref{skizze}). The motors can perform three different actions: if they are bound to the filament, which is represented as a one-dimensional lattice, they can make a step of size $d$ or detach from the filament. Both events occur with force-dependent rates $s(F_i)$ and $k^\pm_d(F_i)$, respectively. Once a motor has detached from the filament, it can reattach  in the interval $\pm L_0$ around the cargo's center of mass with a force-independent rate $k_a$. Due to the de-/attachment events the number of plus (minus) motors bound to the filament is in the range $0 \leq n_+ \leq N_+$ ($0 \leq n_- \leq N_-$). The resulting force on the cargo at position $x_C(t)$ at time $t$ is then given by the sum of all single forces applied by bound motors
\begin{flalign}
&F (x_C(t),\{x_i\}) =  \sum_{i=1}^{n_++n_-}  F_i (x_i-x_C(t)) 
\end{flalign}
with
\begin{flalign}
& F_i (x_i -x_C(t))  \\ \nonumber
& =\begin{cases}
\alpha (x_i-x_C(t) +L_0), \ \ \  \ &  x_i-x_C(t)<-L_0\\
0 , & |x_i-x_C(t)|<L_0\\
\alpha (x_i-x_C(t)-L_0),  &x_i-x_C(t)>L_0
\end{cases} \nonumber
\end{flalign}
The motor's stepping behavior depends on the surrounding ATP concentration ([ATP]) as well as the force $F_i$ applied to the cargo.
As mentioned above, forces that are acting on the motors change their detachment and stepping rate.  
If the force is acting against the preferred direction the motor may slow down, stop or even invert its direction. The 
maximal force (in absolute value) under which a
motor still walks in its preferred direction is given by $F_S$. 
We shall thus distinguish two regimes for the
stepping behavior, depending on whether the applied
force is smaller than the stall force (regime I)
or larger (regime II).
In regime I ({+end motors: $0\leq F_i<F_S$ ; -end motors: $-F_S<F_i\leq 0$}), 
we use a 2-state Michaelis-Menten equation which was introduced in \cite{schnitzer_v_b2000} to model kinesin stepping: we describe the motor stepping rate with ATP playing the role of the stepping catalyst. On the one hand we have ATP which is needed to convert the chemical energy in mechanical energy to perform a step, on the other hand we have the force which acts on the motor and reduces its stepping rate. So we use
\begin{equation}\label{MMe}
s(F_i,[ATP]) =\frac{ k_\text{cat}(F_i)[ATP]}{[ATP]+k_\text{cat}(F_i)\cdot k_\text{b}(F_i)^{-1}},
\end{equation}
where $k_\text{cat}$ is the rate constant for the ATP \underline{cat}alysis, $k_\text{b}$ is a second-order rate constant for ATP \underline{b}inding. Both rate constants depend on the load force. As the stepping rate decreases with increasing absolute force $|F_i|$, $k_\text{b}(|F_i|)$ declines faster with force than $k_\text{cat}(|F_i|)$. Schnitzer \textit{et al.} \cite{schnitzer_v_b2000} introduce a Boltzmann-type force relation for the rate constants
\begin{equation}
k_j(F_i) = \frac{k_j^0}{p_j + q_j \exp(\beta F_i\Delta)} \ \ \ \ \ \ j=\{\text{cat, b} \}
\end{equation} 
with a constant $k_j$, $p_j + q_j = 1$, $\beta = (k_\text{b} T)^{-1}$ and $\Delta $ which is the characteristic distance over which the load acts. It was measured for kinesin \cite{schnitzer_v_b2000} and dynein \cite{toba2006} that the stepping rate, depending on [ATP] and the load force $F_i$, can be described by eq. (\ref{MMe}). The ATP-concentration can be varied considerably in \textit{in vitro} experiments while  \textit{in vivo} processes other than the motors' stepping rates would be influenced as well. 

It is not conclusively clarified whether the stall force of kinesin and dynein depends on [ATP] or not. Following the experimental results of \cite{mallik_g2004} we choose dynein's stall force changing with [ATP] in an affine linear manner from $F_S([ATP]=0$ mM) $= 0.3$ pN to $F_S([ATP]=1$ mM)$=1.2$ pN where it saturates, while we leave kinesin's stall force constant at 2.6 pN. This determines $\Delta$ as given in in Table~\ref{paraset} to ensure that the stepping rate is zero at stall. We find $\Delta$ to get $s(F_S)\ll1$ by solving the equation
\begin{equation}\label{stall}
\frac{ k_\text{cat}(F_S)[ATP]}{[ATP]+k_\text{cat}(F_S)\cdot k_\text{b}(F_S)^{-1}} =  10^{-13} \text{ s}^{-1},
\end{equation} 
where we use $k_\text{cat}^0$, $k_\text{b}^0$, $q_\text{cat}$ and $q_\text{b}$ from Schnitzer \textit{et al.} \cite{schnitzer_v_b2000} and given in Table \ref{paraset} to define the stepping rate
$s(|F_i|,[ATP])$.
In regime II (+end motors: $F_i\geq F_S$ ; -end motors: $-F_S\geq F_i $), 
the motors can walk backwards with the constant rate
\begin{align}
s(F_i) = \frac{v_b}{d},
\end{align}
corresponding to the backward velocity $v_b$, which is at least one order smaller than the force-free forward velocity. In the model we neglect any motor position exclusion on the filament since motors can walk on several lanes of the microtubule. Furthermore the MT network is very dense in many cell regions, as for example in axons where the distance between two microtubules is about 20 nm \cite{hirokawa_t2005}. There, motors can walk on multiple microtubules around the pulled cargo.  
We choose the detachment behavior according to \cite{kunwar2011}
\begin{align}
k_d^+(F_i) = \begin{cases}
k_d^0 \exp\left(\frac{|F_i|}{2.5f}\right) &F_i<F_S \\
k_d^0 \left(0.186\frac{|F_i|}{f} + 1.535\right) &F_i \geq F_S 
\end{cases}
\end{align} 
for kinesin and
\begin{equation}
k_d^-(F_i) = \begin{cases}
k_d^0 \exp\left(\frac{|F_i|}{2.5f}\right) \! &F_i>-F_S \\
k_d^0 \left[1.5\left(1-\exp\left(\frac{-|F_i|}{1.97f}\right)\right)\right]^{-1} \! &F_i\leq -F_S 
\end{cases}
\end{equation}
for dynein. We use the force-free detachment rate $k_d^0$ and a standardization force $f=1$ pN which determines the force scale. Note that the behavior of both types of motors differs above stall, where we have a catch-bond like rate (decreasing with $F_i$) for dynein. 
We propagate the system by means of Gillespie's algorithm for time-dependent rates \cite{gillespie1978}, since the force exerted on each motor (and thus the rates) is time-dependent. The viscosity $\eta$ of the cytoskeleton is taken into account when we solve the equation of motion for the cargo
\begin{align}
m \frac{\partial^2 x_C(t) }{\partial t^2 } = -\beta \frac{\partial x_C(t)}{\partial t} + F(x_C(t),\{x_i\}) ,
\label{eqofm}
\end{align}
with Stokes' law $\beta = 6 \pi \eta R$ ($R$: cargo radius, $m$: cargo's mass).
\section{Results and Discussion}\label{results}
\begin{figure}[tb]
\centering
\includegraphics[width=0.4\textwidth]{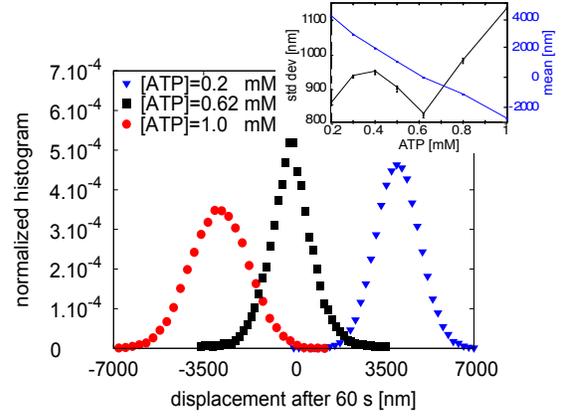}
\caption{Normalized histogram of the displacement after 60 s of cargo propagation. The red circles correspond to [ATP]$=0.2$ mM, the black squares to [ATP]$= 0.62$ mM and the blue triangles to [ATP]$=1.0$ mM. We get for the mean displacement $\mu_{0.2} =4028.3 \pm 1.14 $ nm and {the standard deviation} $\sigma_{0.2} =855.9 \pm 1.13 $ nm for [ATP] = 0.2 mM, $\mu_{1.0} = -2887.5 \pm 3.17 $ nm and $\sigma_{1.0} = 1139.9 \pm 2.59$ nm for [ATP] = 1.0 mM and $\mu_{0.62} = -27.8 \pm 14.08$ nm and $\sigma_{0.62} = 836.9 \pm 11.5$ nm for [ATP] = 0.62 mM. Obviously, the displacement changes direction from plus-end bias for stronger plus motors at low [ATP] to minus-end bias at saturating [ATP]. The inset shows the non-monotonous curve of the standard deviation of the displacement (black) against the ATP concentration while the mean displacement (blue) decreases monotonously. }
\label{drift}
\end{figure}
First we simulate the model in order to measure the distribution of the cargo
displacement for one-minute intervals (see Fig. \ref{drift}). The results, which we obtained for 
the  model's parameters given in 
Table \ref{paraset}, show that we can tune the mean velocity (or bias) of the 
transported cargo continuously by means of the ATP-concentration. 
Remarkably, the cargo {\it slows down with increasing} ATP-concentration until 
the bias vanishes (at an ATP concentration
of $0.62$ mM for our choice of the parameters). For larger concentrations of ATP the bias is inverted
and its absolute value increases again. In contrast to scenarios based on signaling processes where a protein would up- or down-regulate only one motor species, here the inversion of the bias results from the difference in the [ATP] dependence of both motors' activities. In particular dynein detaches more easily with decreasing ATP concentration and so the bias is then directed to the plus-end. We also notice that an unbiased motion of the cargo can only be obtained in a rather narrow interval of ATP-concentration, so that biased motion is generic. 
\begin{figure}[tb]
\centering
\includegraphics[width=0.35\textwidth]{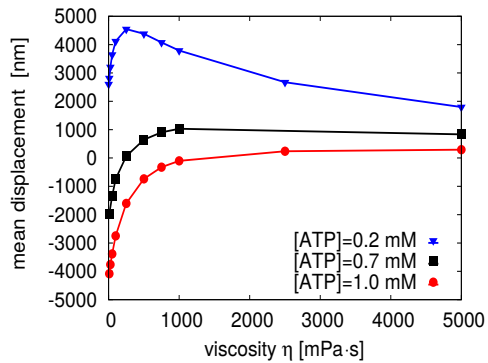}
\caption{Mean displacement after 60 s of cargo propagation for different cytoskeleton viscosities $\eta$. The red circles corresponds to the [ATP]$=0.2$ mM, the black squares to [ATP]$=0.7$ mM and the blue triangles to [ATP]$=1.0$ mM. For small [ATP] the mean displacement-viscosity relation shows a non-monotonous behavior. In the case of [ATP]=0.7 mM we observe a change in direction with increasing viscosity.}
\label{mdvisc}
\end{figure}
The step rate of the motors depend not only on the ATP concentration but also on the force which is applied to the motor. The external force applied to the moving cargo can be parametrized via the viscosity of the environment. In Fig. \ref{mdvisc} we show the dependence of the mean displacement on the 
viscosity (for different ATP concentrations). We observe a non-trivial dependence of the bias on the viscosity. Counterintuitively, the absolute value of the bias can increase with increasing viscosity. For intermediate  ATP concentrations (see, e.g., 0.7 mM in  Fig. \ref{mdvisc} ) one observes that the cargo is changing its direction with increasing viscosity. This effect can be used in order to leave crowded areas in the cellular environment,  which correspond to high effective viscosities,  and enhance thereby the efficiency of the motor-driven transport. Here we find the bias-inversion regime for a limited range of [ATP]. Depending on the cargo and the transportation task to be fulfilled it could be more favorable to be in this bias-inversion regime or in a regime of strong bias. Evolution could have selected the regime by adjusting some other parameters like the motors concentration or an asymmetry in the motor properties. Indeed, a change in $N_+, \ N_-$ modifies the values of the external control parameters at which we observe the reversal of the bias.
\begin{figure}[tb]
\centering
\includegraphics[width=0.4\textwidth]{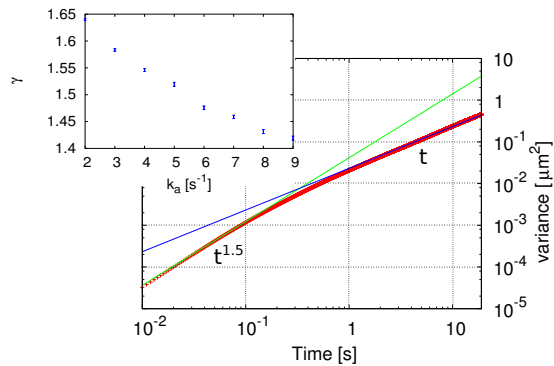}
\caption{{Variance of the numerically generated
trajectories with [ATP]=1 mM. One observes at 0.5 s a
crossover from enhanced diffusion going with $t^{\gamma}, \gamma=1.5$ (green line)
to normal diffusion proportional to $t$ with a diffusion coefficient $D=0.04 \mu\text{m}^2 \text{s}^{-1}$ (blue line). The inset
shows how the exponent varies with the attachment rate $k_a$.}}
\label{visc}
\end{figure}
The cargo-mediated coupling of the engaged motors induces a time-correlated motion of the cargo. These correlations lead to non-gaussian displacement distributions  of 
the cargo at short finite time intervals (see Fig. \ref{drift}). The range of the correlations can by estimated from the functional behavior of the cargo's variance 

\begin{equation}
\Var[x_C(t)] = \langle x_C(t)^2 \rangle - \langle x_C(t) \rangle^2.
\end{equation} 
Unlike in most experiments we chose the variance instead of the mean square displacement. This enables us to disentangle correlation effects from non-diffusive behavior that originates from the bias of the motor-cargo complex. Thus by measuring the variance we are able to analyze if our model exhibits a persistent motion with a (finite) correlation time. Indeed, our simulation results show a transition from enhanced diffusion at short times to normal diffusion. For the set of parameters of Table \ref{paraset} we observe that the variance is growing as $t^{1.5}$ {at short time scales}. Thus, our model generates a superdiffusive behavior which indicates cooperation between motors induced by cargo-mediated coupling. Our results also show that the exponent is not universal {(inset Fig. \ref{visc})}. By varying the attachment rate, which depends strongly on the cellular environment, we have been able to observe exponents between 1.4 and 1.6, so in the same range as observed experimentally \cite{kulic2008,caspi_g_e2002}. 

Asymptotically we observe purely (biased or unbiased) diffusive behavior. This asymptotic diffusive regime is generic for our model as long as we assume cargo 
motion on a single filament. By contrast, on a branched microtubule network the asymptotic behavior may be influenced by the structure of the 
network as shown in the case of the microtubule-network of giant fibroblast \cite{caspi_g_e2002}, where sub-diffusive behavior has been observed.

\begin{figure}[tb]
\centering\begingroup%
  \makeatletter%
  \providecommand\color[2][]{%
    \errmessage{(Inkscape) Color is used for the text in Inkscape, but the package 'color.sty' is not loaded}%
    \renewcommand\color[2][]{}%
  }%
  \providecommand\transparent[1]{%
    \errmessage{(Inkscape) Transparency is used (non-zero) for the text in Inkscape, but the package 'transparent.sty' is not loaded}%
    \renewcommand\transparent[1]{}%
  }%
  \providecommand\rotatebox[2]{#2}%
  \ifx\svgwidth\undefined%
    \setlength{\unitlength}{226.15584432bp}%
    \ifx\svgscale\undefined%
      \relax%
    \else%
      \setlength{\unitlength}{\unitlength * \real{\svgscale}}%
    \fi%
  \else%
    \setlength{\unitlength}{\svgwidth}%
  \fi%
  \global\let\svgwidth\undefined%
  \global\let\svgscale\undefined%
  \makeatother%
  \begin{picture}(1,0.58699732)%
    \put(0,0){\includegraphics[width=\unitlength]{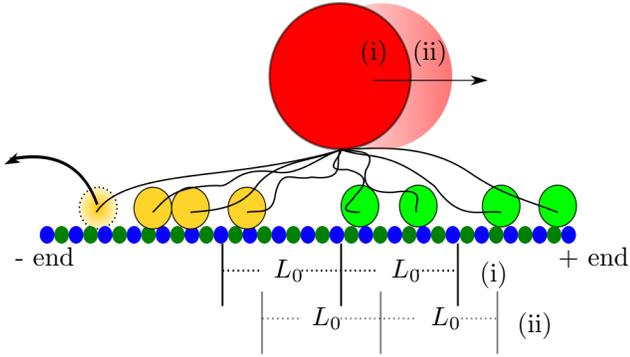}}%
    \put(0.92270266,0.15041663){\color[rgb]{0,0,0}\makebox(0,0)[lb]{\smash{+ end}}}%
    \put(0.01746945,0.15041663){\color[rgb]{0,0,0}\makebox(0,0)[lb]{\smash{- end}}}%
    \put(0.44803743,0.16075078){\color[rgb]{0,0,0}\makebox(0,0)[lt]{\begin{minipage}{0.05996617\unitlength}\raggedright $L_0$\end{minipage}}}%
    \put(0.64411622,0.16142699){\color[rgb]{0,0,0}\makebox(0,0)[lt]{\begin{minipage}{0.0661397\unitlength}\raggedright $L_0$\end{minipage}}}%
    \put(0.71014211,0.08146725){\color[rgb]{0,0,0}\makebox(0,0)[lt]{\begin{minipage}{0.0661397\unitlength}\raggedright $L_0$\end{minipage}}}%
    \put(0.51206442,0.08146725){\color[rgb]{0,0,0}\makebox(0,0)[lt]{\begin{minipage}{0.0661397\unitlength}\raggedright $L_0$\end{minipage}}}%
    \put(0.59194442,0.49422313){\color[rgb]{0,0,0}\makebox(0,0)[lb]{\smash{(i)}}}%
    \put(0.67898601,0.49422313){\color[rgb]{0,0,0}\makebox(0,0)[lb]{\smash{(ii)}}}%
    \put(0.79147097,0.12371898){\color[rgb]{0,0,0}\makebox(0,0)[lb]{\smash{(i)}}}%
    \put(0.85408052,0.04064423){\color[rgb]{0,0,0}\makebox(0,0)[lb]{\smash{(ii)}}}%
  \end{picture}%
\endgroup%
\caption{Sketch of the motor \textit{reservoir}. Assuming that four motors of both kinds are bound to the filament in the given configuration, three minus-end directed motors (yellow) and two plus-end directed motors (green) pull the cargo. If the leftmost motor detaches, the cargo will move to the right and therewith also the area of untensioned motors $x_C(t) \pm L_0$ (going from state (i) to (ii)). In the new motor configuration still three minus-end directed motors and two plus-end directed motors exert a force on the cargo.} \label{res}
\end{figure}

\subsection{Influence of the model parameter}
We want to analyze the influence of the model parameters on the system's dynamics more in detail. 
We shall now vary one single parameter at a time. The
other parameters will keep their value of Table \ref{paraset},
while two values for $[ATP]$ will be considered ($0.2$ and $1$
mM) corresponding to a positive and negative
cargo bias, respectively.

First we vary the total number of motors while keeping
$N_+=N_-$. We observe that the cargo slows down with increasing motor number and the absolute value
of the average bias decreases. One can explain this effect by taking a close look to the motor configuration.

Motors always attach to the microtubule in the force-free area $x_C(t) \pm L_0$, and it takes some time
before they can apply a force on the cargo.
The set of these non-pulling attached motors can be seen as
a {\em reservoir} of motors that can be mobilised very quickly
when the configuration of pulling motors changes, as illustrated
in Fig. ~\ref{res}: when a pulling motor detaches, the cargo
moves and new motors from the {\em reservoir} are involved in
the tug-of-war. During this process, the more motors there are in the
{\em reservoir}, the more limited the displacement of the cargo is.

The above argument involves only motors attached to the microtubule. Another way to increase this number is to increase the attachment rate $k_a$. Then the bias should decrease.
Conversely, increasing the \textit{detachment rate} $k_d^0$
should lower the number of attached motors and thus
increase the bias. This is what is indeed observed in
our simulations.

Eventually, if we increase the \textit{force-free velocity}
$v_f$, the absolute value of the bias increases for
both $[ATP]$ values.
Indeed, motors will then exit the force-free area more
rapidly, depleting the {\em reservoir} of non-pulling attached
motors.

\section{Conclusion}

In this work we study the effect of environmental features on cargos transported by teams of motors.

We find that trajectories of such motor-cargo complexes exhibit different dynamic regimes. Generically one observes a biased motion of 
the cargo, since the activity of dynein and kinesin motors can only be balanced in a very narrow interval of the external control parameters (here
viscosity and ATP-concentration). 
The mechanical coupling of the motors induces time-correlated cargo trajectories. In the comoving frame, the correlations lead to super-diffusive behavior 
at short times. The exponents which describe the time evolution of the variance at short times, as well as the time at which one observes the transition to gaussian displacement distributions are parameter-dependent. For typical parameter combinations the crossover to diffusion takes place at time intervals of the order of one second. Here we have not considered thermal fluctuations. On-going work shows that thermal noise is only relevant for time scales even smaller than those of Fig. \ref{visc} and does not alter the superdiffusion.

Naively one would expect faster cargo transport for higher concentrations of ATP and lower viscosities. In this work we show that this is not the case. The response to both environmental parameters is non-monotonous. 
Counterintuitively, the cargo may accelerate with increasing viscosity for a given parameter-range. We also find an unexpected response of the cargo dynamics to variation of the ATP concentration. We not only observe a change of the bias but also a non-monotonous dependence of the width of the displacement distribution on the ATP concentration. The recent progress in dynein motility assays allows now to test our predictions \textit{in vitro}. The realization of such \textit{in vitro} assays would allow to study the mechanic coupling of molecular motors via the cargo in a much simpler environment as in the living cell. 
\textit{In vivo} we do not expect relevant variations
of ATP concentration.
However, the ATP hydrolysis could be locally
regulated at the level of motors.

In crowded areas of the cell, interactions between different cargos as well as between cargo and filament may apply forces on the molecular motors. 
Our model results show that, since kinesin and dynein motors respond differently to external forces, 
this may lead to an inversion of the cargo's bias, even if no distinct control mechanism is applied.  
The ability of the cargo to change its direction in a crowded environment, i.e. an environment of higher 
effective viscosity, may give an argument why the transport of cargo by oppositely directed motors
can be beneficial for the cell. 
\begin{acknowledgments}
This work was supported by the Deutsche 
Forschungsgemeinschaft (SFB1027,GRK1276) and the BMBF (FKZ 03X0100C).
\end{acknowledgments}
\begin{table*}[htb]
\begin{center}
\begin{tabular}{|c|c|c|c|} 
\hline
 & \textbf{kinesin} & \textbf{dynein} & Ref.\\ \hline \hline
 $d$ & \multicolumn{2}{c|}{8 nm}  & \cite{Carter,toba2006}\\ \hline
 $N_\pm$ &  \multicolumn{2}{c|}{5} & \cite{Welte1998} \\ \hline
 $v_{f}$ & \multicolumn{2}{c|}{$1000 \ \textnormal{nm/s} $} & \cite{Carter,toba2006}  \\ \hline
$v_{b}$ & \multicolumn{2}{c|}{$6 \ \textnormal{nm/s}$} & \cite{Carter,Gennerich}$^*$\\ \hline
$\alpha$ &  \multicolumn{2}{c|}{0.1 $\textnormal{pN/nm}$}& \cite{kunwar2011}$^*$\\ \hline
 $k_a$ & \multicolumn{2}{c|}{${5 \ }{\text{s}}^{-1}$}& \cite{mueller_k_l2008,leduc2004} \\ \hline
 $k_d^0$ & \multicolumn{2}{c|}{$5\ {\text{s}}^{-1}$} & \cite{kunwar2011}$^*$\ \\ \hline
  $f$ & \multicolumn{2}{c|}{1 pN} & \\ \hline
$F_S$ & 2.6 pN & 0.3-1.2 pN &\cite{Mallik2004,Shubeita} \\ \hline
 $k_d(F)$ & $\begin{cases}
  k_d^0\exp\left(\frac{|F|}{F_D}\right),  &  F<F_S \\
  k_d^0(1.535+0.186\cdot \frac{|F|}{f}), & F\geq F_S \end{cases}$ 
  & $\begin{cases}
  k_d^0\exp\left(\frac{|F|}{F_D}\right),  &  F>-F_S \\
  k_d^0\left[1.5\left(1-\exp\left(\frac{-|F|}{1.97f}\right)\right)\right]^{-1}, &  F\leq -F_S \end{cases}$  & \cite{kunwar2011}$^*$\\ \hline
  $k_\text{cat}^0$ &   \multicolumn{2}{c|}{$v_f\cdot d^{-1}$} & \cite{schnitzer_v_b2000} \\ \hline
  $k_\text{b}^0$ &   \multicolumn{2}{c|}{1.3 $\mu$M$^{-1}$s$^{-1}$}& \cite{schnitzer_v_b2000}\\ \hline
  $q_\text{cat}$ &  \multicolumn{2}{c|}{$6.2\cdot 10^{-3}$}& \cite{schnitzer_v_b2000} \\ \hline
  $ q_\text{b}$ & \multicolumn{2}{c|}{$4\cdot 10^{-2}$} & \cite{schnitzer_v_b2000}\\ \hline
$\Delta$ &4267.3 nm & $\max\left(\frac{2.9\cdot 10^5}{[ATP]^{0.3}}-28111.1,8534.6\right)$ nm & eq. (\ref{stall})\\ \hline  \hline
\multicolumn{4}{|c|}{Environment} \\ \hline \hline
$\eta$ &  \multicolumn{2}{c|}{ 100 mPa$\cdot$s} & \cite{kulic2008}$^*$\\ \hline \hline
\multicolumn{4}{|c|}{Cargo} \\ \hline \hline
$R$ &  \multicolumn{2}{c|}{$250$ nm }& \cite{thiam2013}$^*$ \\ \hline
$m$ &  \multicolumn{2}{c|}{$5\cdot 10^{-19}$-$5\cdot 10^{-17}$ kg} &\\ \hline
\end{tabular}
\caption{The second and third column show the simulation parameters for kinesin and dynein, respectively. The fourth column gives the references providing experimental basis to these values. The $^*$ indicates that the experimental support gives the order of magnitude only.}
\label{paraset}
\end{center}
\end{table*}

\bibliographystyle{eplbib}
\bibliography{biblio.bib}

\end{document}